\documentclass[letterpaper, 10 pt, conference]{ieeeconf}  

\IEEEoverridecommandlockouts                              

\usepackage{graphicx}
\usepackage{float}
\usepackage{amsmath}
\usepackage{subfiles}
\usepackage{enumerate}
\usepackage[T1]{fontenc}
\usepackage{amssymb}
\usepackage{subfig}
\usepackage{color}
\usepackage{soul}
\usepackage[noadjust]{cite}

\renewcommand{\baselinestretch}{0.966}

\title{\LARGE \bf Frequency Superposition -- A Multi-Frequency Stimulation Method in SSVEP-based BCIs}

\author{Jing Mu, David B. Grayden, Ying Tan, and Denny Oetomo%
\thanks{This work was supported by the Valma Angliss Trust.}
\thanks{J. Mu, Y. Tan, and D. Oetomo are with the Department of Mechanical Engineering, The University of Melbourne, Parkville, VIC 3010, Australia.}
\thanks{D. B. Grayden is with the Department of Biomedical Engineering, The University of Melbourne, Parkville, VIC 3010, Australia.}
\thanks{{\tt\small j.mu@student.unimelb.edu.au, \{grayden, yingt, doetomo\}@unimelb.edu.au}}
}

\begin{document}

\maketitle
\thispagestyle{empty}
\pagestyle{empty}

\begin{abstract}

The steady-state visual evoked potential (SSVEP) is one of the most widely used modalities in brain-computer interfaces (BCIs) due to its many advantages. However, the existence of harmonics and the limited range of responsive frequencies in SSVEP make it challenging to further expand the number of targets without sacrificing other aspects of the interface or putting additional constraints on the system. This paper introduces a novel multi-frequency stimulation method for SSVEP and investigates its potential to effectively and efficiently increase the number of targets presented. The proposed stimulation method, obtained by the superposition of the stimulation signals at different frequencies, is size-efficient, allows single-step target identification, puts no strict constraints on the usable frequency range, can be suited to self-paced BCIs, and does not require specific light sources.
In addition to the stimulus frequencies and their harmonics, the evoked SSVEP waveforms include frequencies that are integer linear combinations of the stimulus frequencies.
Results of decoding SSVEPs collected from nine subjects using canonical correlation analysis (CCA) with only the frequencies and harmonics as reference, also demonstrate the potential of using such a stimulation paradigm in SSVEP-based BCIs.

\end{abstract}


\section{Introduction}

The steady-state visual evoked potential (SSVEP) is a robust method to induce brain activity detected using electroencephalography (EEG). The advantages of SSVEP include its relatively high information transfer rate, minimum training requirements, and ability to provide larger numbers of targets or classes than other brain-computer interface (BCI) approaches, such as motor imagery \cite{nicolas2012brain}.

Currently, SSVEP-based BCIs are mostly designed using a single frequency as the stimulus for each target \cite{muller2007control,chen2015high,han2018novel}. However, because of the existence of harmonics and the limited range of responsive frequencies in the SSVEP, single-frequency SSVEP performance falls with large numbers of targets, which demands a large number of single frequencies with narrower frequency differences between them. 
Therefore, stimuli that use more than single frequencies have been developed to handle larger numbers of targets \cite{shyu2010dual,jia2011frequency,zhang2012multiple,hwang2013new,kimura2013ssvep,chen2013brain,chang2014amplitude}.

Existing multi-frequency methods can be categorised into four groups -- spatially separated, temporally separated, amplitude modulation, and multi-modal stimulation.
Spatially separated methods use different light sources to show different frequencies, such as using two LEDs as a single stimulation unit \cite{shyu2010dual} or flashing the two alternating areas of a checkerboard with different frequencies \cite{hwang2013new}. While feasible, they usually require a larger area for the presentation of the stimulus, resulting in poor spatial resolution of target display. 
Furthermore, it has been found that SSVEP may get attenuate at the stimulation frequencies when two stimulation units are placed adjacent to each other \cite{mu2019spatial}.
Temporally separated methods combine two frequencies temporally; the visual stimulus is shown as a sequence of frequencies that are presented one after the other \cite{zhang2012multiple,kimura2013ssvep}. This introduces limits on the speed of the BCI.
Amplitude modulation uses a high frequency carrier that is modulated by a lower frequency signal \cite{chang2014amplitude}. It provides higher size and time efficiency compared to the aforementioned methods, but puts constraints on the frequency combinations that can be applied since the carrier frequency needs to be much higher than the frequency of the modulation signal. By size efficient, we mean that it requires only the minimum unit of medium in stimulation, in contrast to the spatially separated methods where multiple units are required.
Multi-modal stimulation uses different modalities to present multiple frequencies or information, such as using both frequency and phase as variables in the stimuli \cite{jia2011frequency} or showing two frequencies using changes in both brightness and colour \cite{chen2013brain}. Compared to the other methods, this approach is naturally the most effective for expanding the number of stimuli with limited frequency options. However, a reliance on frequency and phase coding requires an accurate timing of the cue or trigger for phase identification, hence it is not suited for self-paced BCIs where the subject is supposed to have full control of the interface. Brightness and colour intermodulation introduce higher requirements on the complexity of the visual display unit.

In this paper, a new multi-frequency stimulation method, frequency superposition, is presented, which demonstrates how multiple single-frequency signals can be combined to address the weaknesses of existing methods. Frequency superposition enables the presentation of more stimuli with fewer frequencies compared to single frequency stimulation. Frequency superposition is size efficient, allows single-step target identification, puts no strict constraints on the frequency range, and is suitable for self-paced BCIs. This method does not require specific light sources, and can work with even single-colour light sources.
Human-subject performance using this method is gathered, analysed and decoded. The features in the resulting SSVEP are explained and the decoding results with a non-trained decoder are presented as preliminary demonstration of the potential of frequency superposition.

\section{Methods}\label{sec:methods}

\begin{figure}
    \centering
    \vspace{0.25cm}
    \includegraphics[trim={0.2cm 1.5cm 2.3cm 1.2cm}, clip, width=0.9\linewidth, keepaspectratio]{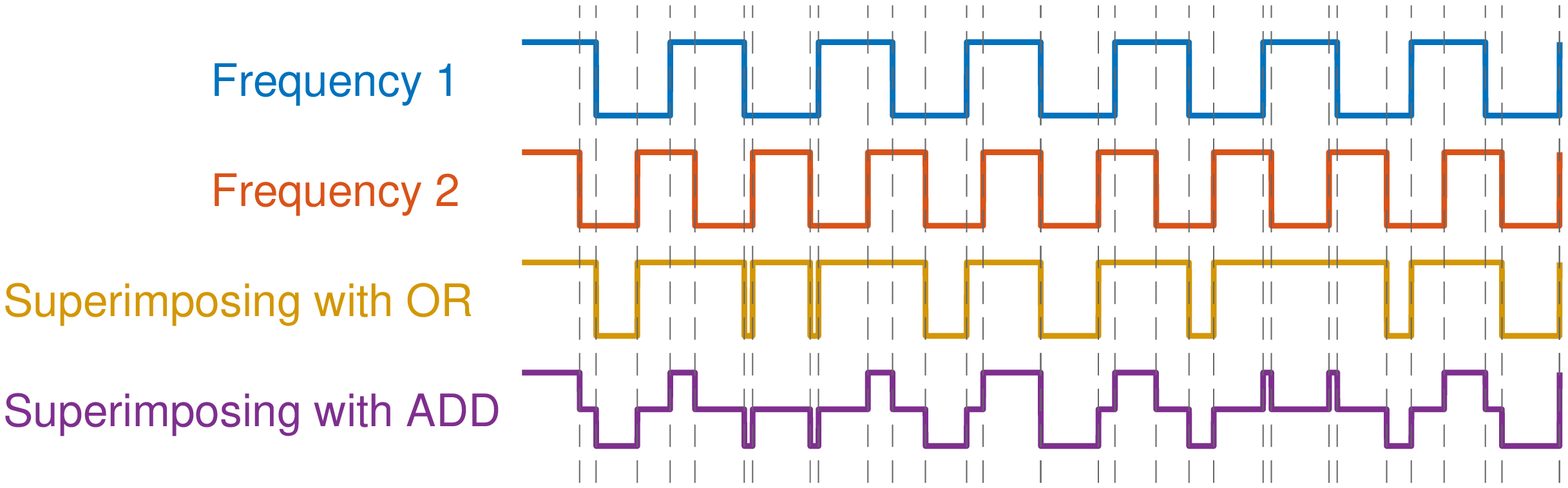}
    \caption{Examples of dual-frequency superposition. Frequency 1 (blue) and Frequency 2 (red) are superimposed using either an OR operation (yellow) or ADD operation (purple).}
    \label{fig:freqSup}
\end{figure}

In frequency superposition, independent signals are first generated at each desired stimulation frequency and superimposed on the stimulation medium. Fig. \ref{fig:freqSup} shows two examples of dual-frequency superposition of square waveforms (blue and red) using ``OR'' and ``ADD'' methods, where ``OR'' indicates that the visual stimulus is set to full brightness if either of the waveforms is HIGH (yellow) and ``ADD'' sums the two waveforms (purple). The technique can also be applied to signals with other waveforms and with more than two frequencies. In this paper, two 50\% duty cycle square waves with the OR operator was selected without loss of generality of the main idea.

\subsection{Experimental Setup}

\subsubsection{Stimulation setup}
\begin{figure}
    \centering
    \includegraphics[width=0.57\linewidth, keepaspectratio]{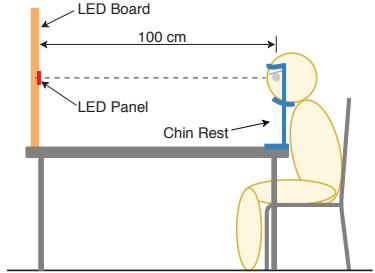}
    \caption{Experimental setup. A chin rest maintains a 100 cm distance between the subject and the LED board and keeps the centre of the LED board at eye level in the sagittal plane of the subject.}\vspace{-0.5cm}
    \label{fig:Experiment}
\end{figure}

A red LED stimulation setup was constructed with Adafruit NeoPixel 8 $\times$ 8 RGB LED panels (70 mm $\times$ 70 mm) mounted on medium-density fibreboard. As shown in Fig. \ref{fig:Experiment}, subjects were positioned 100 cm from the LED board during the experiments with the centre of the LED board at eye level and in the sagittal plane of the subject. A chin rest was used during the experiments to maintain a consistent distance between the subject and the LED board throughout the experiment.

\subsubsection{EEG}
EEG data were recorded from six channels (PO3, POz, PO4, O1, Oz, O2; international 10-20 system) at 512 Hz sampling rate using a g.USBamp amplifier with g.SAHARA dry electrodes (g.tec medical engineering GmbH, Austria). Data were filtered with a 50 Hz notch filter and a 0.5-60 Hz bandpass filter during data acquisition in the g.USBamp settings. Reference and ground electrodes were placed on the left and right mastoids, respectively.

\subsection{Experimental Protocol}

The experiment consisted of presentations of the frequency pairs: 7 \& 9 Hz, 7 \& 11 Hz, 7 \& 13 Hz, 9 \& 11 Hz, 9 \& 13 Hz, and 11 \& 13 Hz.
Each trial took 40 s in total. A trial started with a 10 s visual cue notifying the subject that the trial was about to start. This was then followed by a 30 s stimulation period during which a single dual-frequency visual stimulus was presented, with 20 s rests between trials.

\subsection{Subjects}
Nine healthy subjects (four females and five males aged 22-33 years, mean 26.8 years, standard deviation 3.7) participated in the experiment. All subjects had normal or corrected-to-normal vision and right hand as dominant hand. Subjects 3, 4, 6, 7, and 8 were na\"ive to SSVEP-based BCIs. 

This experiment was approved by the University of Melbourne Human Research Ethics Committee (Ethics ID 1851283). Written consent was received from all subjects prior to the experiments.

\subsection{Data processing for pattern extraction in frequency superposition}\label{sec:dataProcessing}

Thirty seconds of EEG recording of each frequency pair was processed. A spatial filter was first applied by subtracting the average of the measurements across all channels from the measurement from channel Oz. A band-pass filter (1.8-60) Hz was then applied to the spatially filtered data and the fast Fourier transform (FFT) was calculated. The \textsc{Matlab} (MathWorks, R2020a) command \textit{bandpass} was called to band-pass filter with default settings (type of impulse response was selected by \textsc{Matlab}; steepness: 0.85; stopband attenuation: 60).

\begin{figure*}[ht]
    \centering
    \vspace{0.23cm}
    \includegraphics[trim={2.2cm 0.5cm 2.5cm 0.67cm}, clip, width=0.78\textwidth, keepaspectratio]{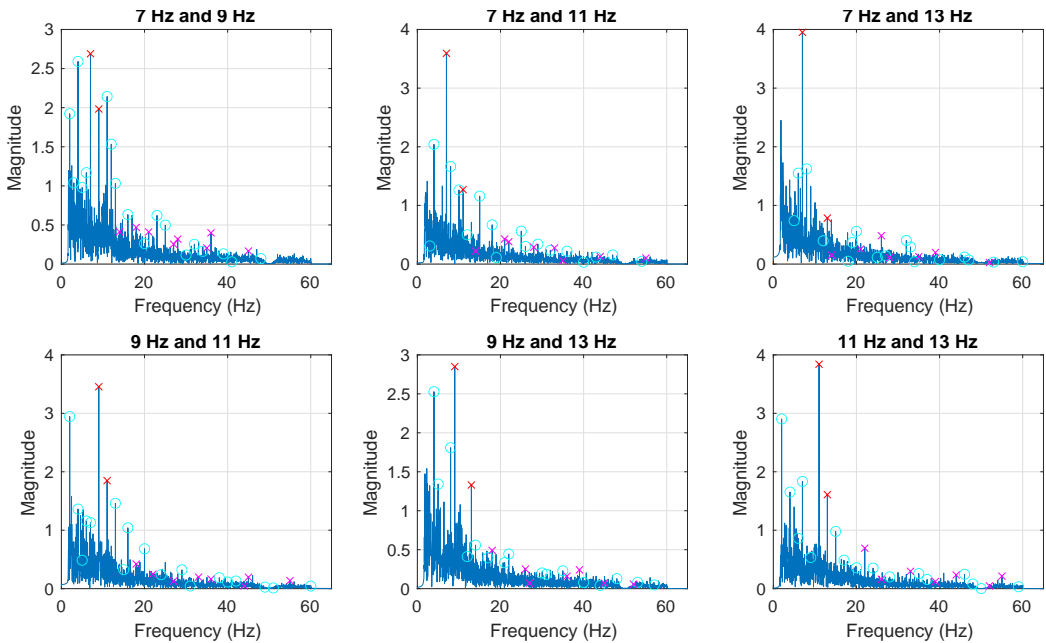}
    \caption{Example FFT plots of SSVEP recordings from Subject 2. Red crosses label the two stimulation frequencies. Magenta crosses label up to the fifth harmonics of the two frequencies. Cyan circles indicate the frequencies that are linear combinations of the two frequencies with the absolute value of coefficients smaller or equal to 3.}\vspace{-0.5cm}
    \label{fig:Result}
\end{figure*}

\subsection{Decoding with Canonical Correlation Analysis (CCA)} 
Canonical Correlation Analysis is a decoding algorithm for SSVEP \cite{lin2006frequency} that compares correlations between the measured multi-channel EEG data $x(t)$ and a reference set $y(t)$. The reference set is designed to contain sinusoidal signals at the fundamental frequencies and harmonics:
\begin{equation}\label{eq:y_single}
    y(t)=
    \begin{bmatrix}
        \sin(2\pi ft)\\
        \cos(2\pi ft)\\
        \vdots \\
        \sin(2\pi N_\mathrm{h} ft)\\
        \cos(2\pi N_\mathrm{h} ft)
    \end{bmatrix},
\end{equation}
where $f$ is the fundamental frequency, $t$ is time, and $N_\mathrm{h}$ is the number of harmonics selected in the reference. CCA requires knowledge of the candidate frequencies to formulate the reference sets. The candidate with the highest correlation with the data is regarded as the decoded output.

Without any prior knowledge of the patterns in the resulting SSVEP with frequency superposition, a standard CCA for single frequency stimulated SSVEP can be extended to multiple frequencies:
\begin{equation}\label{eq:y_multi}
    y_{\textrm{multi}}(t)=
    \begin{bmatrix}
        y_1(t)\\
        y_2(t)\\
        \vdots \\
        y_N(t)
    \end{bmatrix},\ 
    y_i(t)=
    \begin{bmatrix}
        \sin(2\pi f_i t)\\
        \cos(2\pi f_i t)\\
        \vdots \\
        \sin(2\pi N_\mathrm{h} f_i t)\\
        \cos(2\pi N_\mathrm{h} f_i t)
    \end{bmatrix},
\end{equation}
where $f_i$, $i=1,2,\dots,N_\mathrm{f}$ is the $i^{th}$ stimulation frequency and $N_\mathrm{f}$ is the number of frequencies. In this work, SSVEP was stimulated with frequency superposition of two frequencies, so $N_\mathrm{f}=2$.

\section{Results}\label{sec:results}

Fig.~\ref{fig:Result} shows Fast Fourier Transform (FFT) magnitude plots of the recorded SSVEP from Subject 2 as a representative illustration. It can be seen that the peaks consist of the two stimulation frequencies and their harmonics, as well as the linear combinations of the two frequencies with integer coefficients. Take the 7 Hz \& 9 Hz trial as an example: the top six peaks are located at 7, 4, 11, 9, 2, and 12 Hz. The corresponding coefficient pairs for the 7 \& 9 Hz example, listed in Table \ref{tab:linCombCoeff}, are given by
\begin{equation}\label{eq:linComb_7_9}
    f_p = c_1 \times f_1 + c_2 \times f_2,\ f_1 < f_2,
\end{equation}
where $f_p$ is the peak frequency, $c_1$ and $c_2$ are integer coefficients for $f_1$ and $f_2$, respectively. Here, $f_1=7$ Hz, $f_2=9$ Hz.

\begin{table}[h]
    \centering
    \begin{tabular}{|l|c c c c c c|}
        \hline 
        Peak Frequency (Hz) ($f_p$) & 7 & 4 & 11 & 9 & 2 & 12 \\ \hline
        Coefficient of 7 Hz ($c_1$) & 1 & -2 & -1 & 0 & -1 & 3\\
        Coefficient of 9 Hz ($c_2$) & 0 & 2 & 2 & 1 & 1 & -1 \\ \hline
    \end{tabular}
    \caption{The top six peaks $f_p$ with their coefficients for the 7 Hz and 9 Hz frequency pair.}
    \label{tab:linCombCoeff}
\end{table}

Note that the cyan circles in Fig.~\ref{fig:Result} only label linear combinations with both coefficients between $-3$ and $+3$.

\begin{figure}[ht]
    \centering
    \vspace{0.2cm}
    \subfloat[]{
        \centering
        \includegraphics[trim={1.5cm 0.7cm 1.6cm 1.1cm}, clip, width=0.63\linewidth, keepaspectratio]{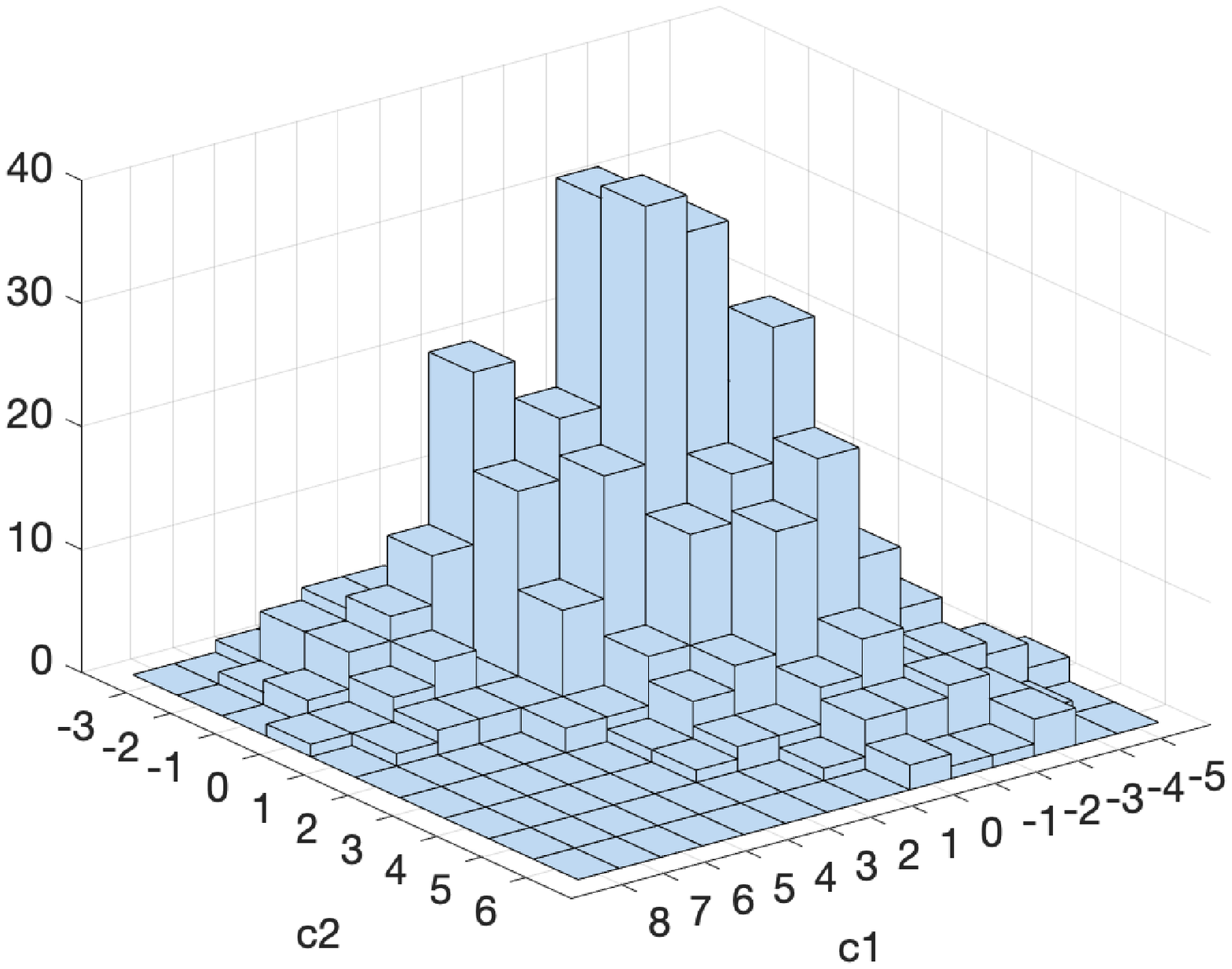}
        \label{fig:order_hist}
    }\\\vspace{-0.15cm}
    \subfloat[]{
        \centering
        \includegraphics[trim={0cm 0cm 0cm 0cm}, clip, width=0.62\linewidth, keepaspectratio]{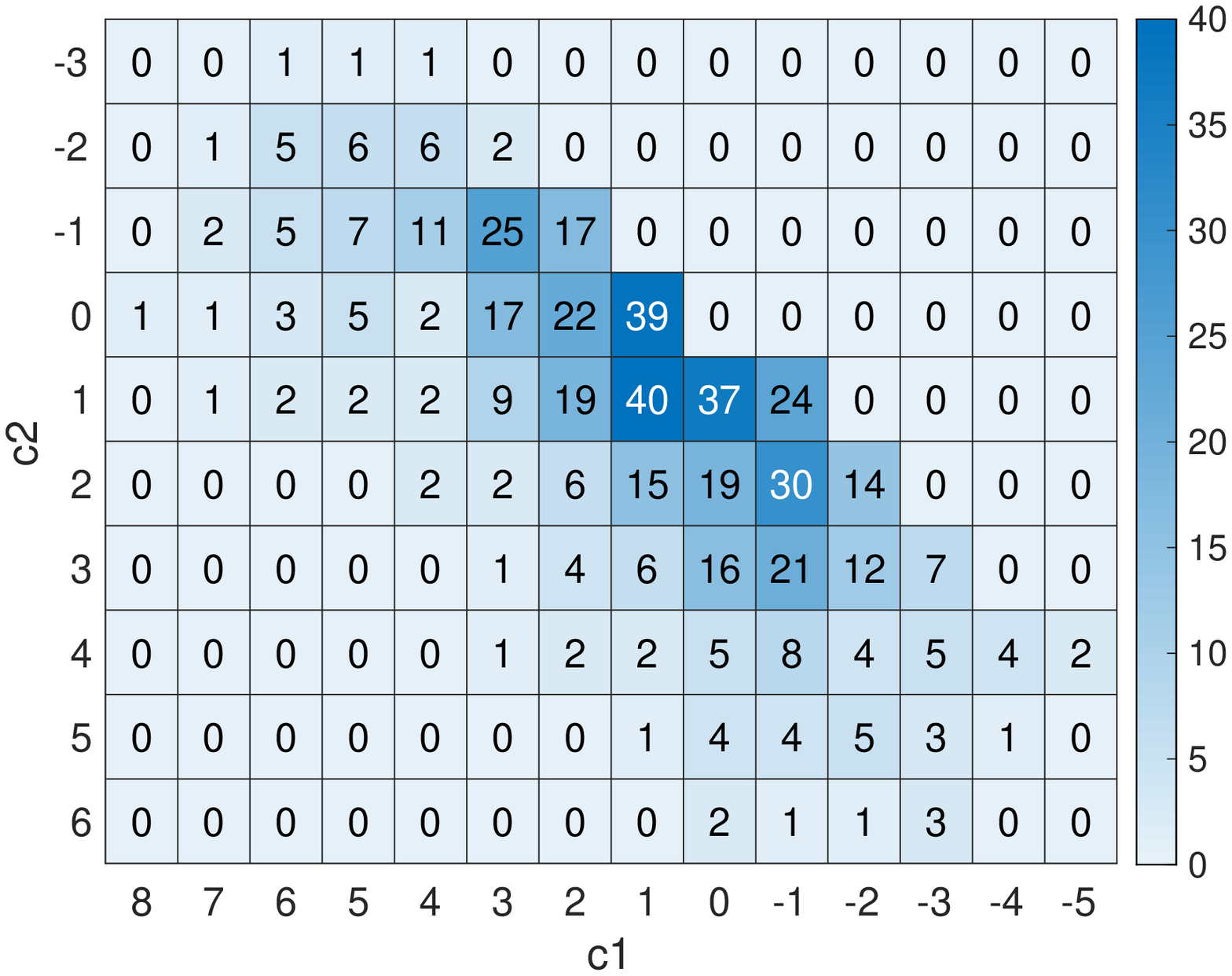}
        \label{fig:order_heatmap}
    }
    \caption{Number of occurrences of each coefficient pair $(c_1, c_2)$. (a) Histogram of the number of appearances of each coefficient pair in the top ten peaks in each trial. (b) The number of counts of each coefficient pair in the top ten peaks in each trial.}\vspace{-0cm}
    \label{fig:order}
\end{figure}

Fig.~\ref{fig:order} shows the number of occurrences of each coefficient pair in the top ten integer peaks in each trial. These peaks were identified by searching for the top peaks located at integer frequencies after calculating FFT of the spatial filtered data. To remove the contribution of background brain activity, the frequency domain signal-to-noise ratio (SNR) was calculated before the peaks were identified. SNR was calculated as the ratio between the magnitude at current frequency and the average magnitude of the 10 neighbouring frequencies (five on each side). The peaks were searched within the frequency range 0.5-60 Hz.

\begin{figure}[ht]
    \centering
    \includegraphics[trim={0.2cm 0cm 1.1cm 0.9cm}, clip, width=0.58\linewidth, keepaspectratio]{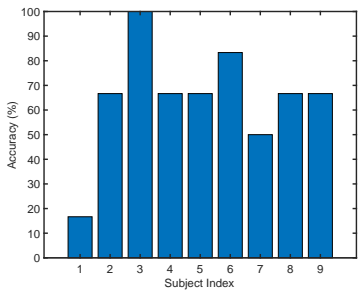}
    \caption{Accuracy for each subject using full 30~s data with CCA decoder when $N_\mathrm{h}=3$.}\vspace{-0.5cm}
    \label{fig:accuracy}
\end{figure}

In order to test if the frequency superposition SSVEP stimuli can be detected with decoding algorithms, CCA for two input frequencies was formulated with the reference signal shown in Eq.\eqref{eq:y_multi}. This was done before analysing the patterns in the resulting SSVEP and with the assumption that the two input frequencies and their harmonics would be part of the pattern.
$N_\mathrm{h}=3$ was selected because the third harmonics could still be observed in the peaks and were not too high in frequency to be substantially attenuated. Decoding accuracy for each subject using the full 30~s recordings is shown in Fig.~\ref{fig:accuracy}.

\section{Discussion}\label{sec:discussion}

Compared to the conventional single frequency stimulation, frequency superposition triggers a more complex response in SSVEP. It contains highly redundant information in the frequency domain with integer linear combinations of the two frequencies. It was observed that the linear combinations follow a pattern where coefficients with smaller absolute value have a higher chance of being observed in the frequency domain of the SSVEP response, especially those that have a smaller combined absolute value.

\subsection{Detecting Multi-Frequency SSVEP with CCA}

CCA was used to decode the experimental data to verify if SSVEP stimulated by frequency superposition could be detected with the decoding algorithm. For this purpose, CCA was formulated without taking into account the actual patterns in the resulting SSVEP, thus only the input frequencies and their harmonics were included in the reference set (Eq.\eqref{eq:y_multi}).
This result does not represent the true potential of frequency superposition stimulated SSVEP, as the other features, such as the linear combinations of the input frequencies, were not exploited in this decoding algorithm.

Fig.~\ref{fig:accuracy} shows the resulting decoding accuracy.
Subject 1 appears to be an outlier and requires further investigation.  Other subjects demonstrated an average of 70.83\% accuracy on frequency superpositioned stimulation. It did not reach the typical SSVEP benchmark accuracy of 80\%; however, the main purpose of this preliminary experiment is to evaluate the viability of the frequency superposition stimulation method.
Improvements to the performance can be gained with using the known SSVEP patterns in decoders and the details in the display arrangements \cite{mu2020comparison}, to bring it to a fair comparison to the conventional performance of single frequency methods. A dry electrode system was used in this experiment, which has been found to yield relatively lower accuracy compared to the wet electrodes \cite{zhu2021open}.

\subsection{Limitations}

It was observed that the resulting SSVEP pattern is different for different subjects when exposed to the same frequency pairs. The highest peaks were not always located at the input frequencies, and the linear combinations observed as the highest peaks are different for different subjects. These introduces more challenges to the use of frequency superposition in SSVEP.

\section{Conclusion}\label{sec:conclusion}

A novel stimulation method, ``frequency superposition'', is introduced to elicit multi-frequency SSVEP.
It was observed that not only the input frequencies and their harmonics could be observed in frequency domain, but also the linear combinations of the input frequencies with integer coefficients.
The experimental results obtained without exploiting properties of frequency superposition stimulated SSVEP showed a 70\% accuracy, lower than the 80\% benchmark for a typical single frequency SSVEP. Further work will exploit the unique properties of the available peaks at linear combination of stimulation frequency.



\bibliographystyle{IEEEtran}
\bibliography{IEEEabrv,ref}

\begin{thebibliography}{10}
\providecommand{\url}[1]{#1}
\csname url@samestyle\endcsname
\providecommand{\newblock}{\relax}
\providecommand{\bibinfo}[2]{#2}
\providecommand{\BIBentrySTDinterwordspacing}{\spaceskip=0pt\relax}
\providecommand{\BIBentryALTinterwordstretchfactor}{4}
\providecommand{\BIBentryALTinterwordspacing}{\spaceskip=\fontdimen2\font plus
\BIBentryALTinterwordstretchfactor\fontdimen3\font minus
  \fontdimen4\font\relax}
\providecommand{\BIBforeignlanguage}[2]{{%
\expandafter\ifx\csname l@#1\endcsname\relax
\typeout{** WARNING: IEEEtran.bst: No hyphenation pattern has been}%
\typeout{** loaded for the language `#1'. Using the pattern for}%
\typeout{** the default language instead.}%
\else
\language=\csname l@#1\endcsname
\fi
#2}}
\providecommand{\BIBdecl}{\relax}
\BIBdecl

\bibitem{nicolas2012brain}
L.~F. Nicolas-Alonso and J.~Gomez-Gil, ``{Brain computer interfaces, a
  review},'' \emph{Sensors}, vol.~12, no.~2, pp. 1211--1279, 2012.

\bibitem{muller2007control}
G.~R. Muller-Putz and G.~Pfurtscheller, ``{Control of an electrical prosthesis
  with an SSVEP-based BCI},'' \emph{IEEE Trans. Biomed. Eng.}, vol.~55, no.~1,
  pp. 361--364, 2007.

\bibitem{chen2015high}
X.~Chen, Y.~Wang, M.~Nakanishi, X.~Gao, T.-P. Jung, and S.~Gao, ``High-speed
  spelling with a noninvasive brain--computer interface,'' \emph{Proc. Natl.
  Acad. Sci.}, vol. 112, no.~44, pp. E6058--E6067, 2015.

\bibitem{han2018novel}
X.~Han, K.~Lin, S.~Gao, and X.~Gao, ``{A novel system of SSVEP-based
  human--robot coordination},'' \emph{J. Neural Eng.}, vol.~16, no.~1, p.
  016006, 2019.

\bibitem{shyu2010dual}
K.-K. Shyu, P.-L. Lee, Y.-J. Liu, and J.-J. Sie, ``Dual-frequency steady-state
  visual evoked potential for brain computer interface,'' \emph{Neurosci.
  Lett.}, vol. 483, no.~1, pp. 28--31, 2010.

\bibitem{jia2011frequency}
C.~Jia, X.~Gao, B.~Hong, and S.~Gao, ``{Frequency and phase mixed coding in
  SSVEP-based brain--computer interface},'' \emph{IEEE Trans. Biomed. Eng.},
  vol.~58, no.~1, pp. 200--206, 2011.

\bibitem{zhang2012multiple}
Y.~Zhang, P.~Xu, T.~Liu, J.~Hu, R.~Zhang, and D.~Yao, ``{Multiple frequencies
  sequential coding for SSVEP-based brain-computer interface},'' \emph{PLoS
  One}, vol.~7, no.~3, p. e29519, 2012.

\bibitem{hwang2013new}
H.-J. Hwang, D.~H. Kim, C.-H. Han, and C.-H. Im, ``{A new dual-frequency
  stimulation method to increase the number of visual stimuli for multi-class
  SSVEP-based brain--computer interface (BCI)},'' \emph{Brain Res.}, vol. 1515,
  pp. 66--77, 2013.

\bibitem{kimura2013ssvep}
Y.~Kimura, T.~Tanaka, H.~Higashi, and N.~Morikawa, ``{SSVEP-based
  brain--computer interfaces using FSK-modulated visual stimuli},'' \emph{IEEE
  Trans. Biomed. Eng.}, vol.~60, no.~10, pp. 2831--2838, 2013.

\bibitem{chen2013brain}
X.~Chen, Z.~Chen, S.~Gao, and X.~Gao, ``Brain--computer interface based on
  intermodulation frequency,'' \emph{J. Neural Eng.}, vol.~10, no.~6, p.
  066009, 2013.

\bibitem{chang2014amplitude}
M.~H. Chang, H.~J. Baek, S.~M. Lee, and K.~S. Park, ``{An amplitude-modulated
  visual stimulation for reducing eye fatigue in SSVEP-based brain--computer
  interfaces},'' \emph{Clin. Neurophysiol.}, vol. 125, no.~7, pp. 1380--1391,
  2014.

\bibitem{mu2019spatial}
J.~Mu, D.~B. Grayden, Y.~Tan, and D.~Oetomo, ``Spatial resolution of visual
  stimuli in {SSVEP}-based brain-computer interface,'' in \emph{9th Int.
  IEEE/EMBS Conf. Neural Eng.}, 2019, pp. 928--932.

\bibitem{lin2006frequency}
Z.~Lin, C.~Zhang, W.~Wu, and X.~Gao, ``{Frequency recognition based on
  canonical correlation analysis for SSVEP-based BCIs},'' \emph{IEEE Trans.
  Biomed. Eng.}, vol.~53, no.~12, pp. 2610--2614, 2006.

\bibitem{mu2020comparison}
J.~Mu, D.~B. Grayden, Y.~Tan, and D.~Oetomo, ``Comparison of steady-state
  visual evoked potential ({SSVEP}) with {LCD} vs. {LED} stimulation,'' in
  \emph{2020 42st Annu. Int. Conf. IEEE Eng. Med. Biol. Soc.}, 2020, pp.
  2946--2949.

\bibitem{zhu2021open}
F.~Zhu, L.~Jiang, G.~Dong, X.~Gao, and Y.~Wang, ``An open dataset for wearable
  ssvep-based brain-computer interfaces,'' \emph{Sensors}, vol.~21, no.~4, p.
  1256, 2021.

\end{thebibliography}

\end{document}